# FlexLMM: a Nextflow linear mixed model framework for GWAS

Saul Pierotti[1], Tomas Fitzgerald[1], Ewan Birney[1]*

[1] European Molecular Biology Laboratory, European Bioinformatics Institute (EMBL-EBI); Cambridge, CB10 1SD, UK
* Corresponding author

## Abstract

**Summary:** Linear mixed models are a commonly used statistical approach in genome-wide association studies when population structure is present. However, naive permutations to empirically estimate the null distribution of a statistic of interest are not appropriate in the presence of population structure, because the samples are not exchangeable with each other. For this reason we developed FlexLMM, a Nextflow pipeline that runs linear mixed models while allowing for flexibility in the definition of the exact statistical model to be used. FlexLMM can also be used to set a significance threshold via permutations, thanks to a two-step process where the population structure is first regressed out, and only then are the permutations performed. We envision this pipeline will be particularly useful for researchers working on multi-parental crosses among inbred lines of model organisms or farm animals and plants.

**Availability and implementation:** The source code and documentation for the FlexLMM is available at https://github.com/birneylab/flexlmm.

## Introduction

Linear Mixed Models (LMMs) are a well-established approach for Genome-Wide Association Studies (GWAS) (Alamin et al., 2022). They can effectively correct for population structure by modelling the covariance of the phenotypes as a function of a genetic relatedness matrix. They can also improve power by implicitly conditioning on loci other than the one being tested, and can be used for estimating the heritability of a phenotype (Yang et al., 2014). Several efficient implementations of LMMs specific for GWAS have been developed. Most notably BOLT-LMM (Loh et al., 2015), fastGWA (Jiang et al., 2019), SAIGE (Zhou et al., 2018), and GCTA (Yang et al., 2011). A different approach but with similar goals is the whole-genome regression implemented in REGENIE (Mbatchou et al., 2021). These software are highly optimised for handling large human datasets such as the UK Biobank (Sudlow et al., 2015). However, this optimisation comes at the price of reduced flexibility in the specification of the statistical models to be compared for evaluating significance. Moreover, in organisms where a community-accepted genome-wide significance threshold is not available, sample permutations are a "gold standard" method for avoiding false positives, but this approach is not valid for LMM if naively implemented (Joo et al., 2016).

We developed FlexLMM as a solution for these two issues. FlexLMM can take in input an arbitrary statistical model for the fixed terms (for example it is possible to modify the genotype encoding to account for dominance), and compares it to an arbitrary null model via a likelihood ratio test. FlexLMM can also natively run permutations. The main issue with permutations in LMMs is the fact that the samples are not exchangeable under the null hypothesis (e.g. the samples are not all equally related to each other). To address this issue,

FlexLMM first estimates the variance-covariance structure from the original datasets, and then regresses it out from the phenotype and design matrix. Only then the genotypes are jointly permuted, preserving the correlation structure across genetic markers and the exchangeability of the samples.

A similar method for performing permutations with LMMs has been recently developed (John et al., 2022), but in this method the permutations are performed on the original dataset where population structure is present. We believe that our approach more accurately tests the null hypothesis that most GWAS users implicitly have - that of a non-zero random effect due to population structure, and zero effect for the specific marker being tested. A permutation approach similar to the one we adopted was also followed in recent work (Scott et al., 2021).

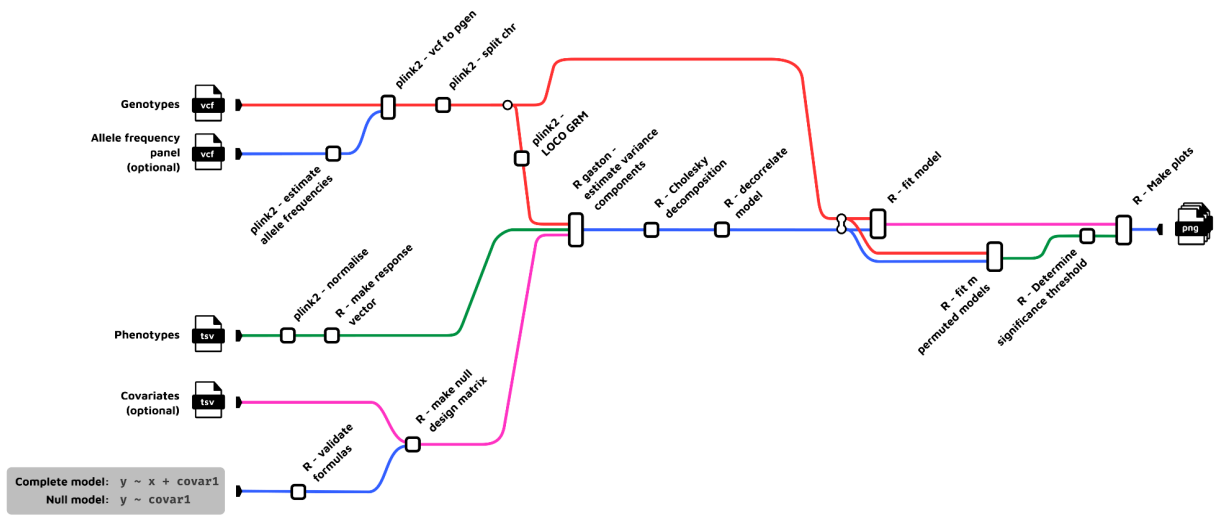

**Figure 1.** Simplified overview of the FlexLMM pipeline. The flow of information is from left to right. Items on the left represent input files or values. Items on the right represent output files. Not all the output and input options and formats are shown.

# Implementation

The FlexLMM logic is coded in the Nextflow Domain Specific Language version 2 (DSL2) (Di Tommaso et al., 2017), while individual steps are mainly taking advantage of the excellent plink2 suite of tools (Chang et al., 2015) and of the R programming language (R Core Team, 2023). Other packages that are used in the pipeline are gaston (Claire & Hervé, 2018), ggplot2 (Wickham, 2016), cowplot (Wilke, 2020), and ComplexHeatmap (Gu et al., 2016). The use of Nextflow enables this pipeline to be transparently run across a wide range of computational platforms with minimal effort. Moreover, we use containerization solutions in all the pipeline steps, allowing for complete reproducibility of computational workflows. For routine genetic data handling (construction of genetic relatedness matrices, format conversions, slicing and parsing genotype matrices) we use the plink2 suite because of its highly optimised code and file formats. For more custom operations, such as the definition of design matrices and contrasts, and model fitting and permutation, we use our own R code for maximum flexibility.

The user is required to provide a VCF file containing the genotypes of interest, and tab-separated files containing sample-specific values for phenotype (optionally also multiple phenotypes to be run in parallel) and covariates. For the moment FlexLMM supports only quantitative phenotypes. The user needs also to provide the statistical model to be used for the GWAS in the form of an R formula, and a null model (also as an R formula) to be used for estimating heritability and for testing significance via a likelihood ratio test.

The pipeline implements a linear mixed model approach to Genome-Wide Association Studies (GWAS), which takes the form:

$$y = Xb + Zu + \varepsilon$$

Where the $n \times 1$ vector of phenotypes $y$ is modelled as a function of a design matrix $X$ of dimensions $n \times q$ of variables treated as fixed effects, that multiplies a $q \times 1$ vector $b$ of fixed effect coefficients. The random effect terms $Zu$ are similarly composed of a $n \times p$ design matrix $Z$ and a random effects vector $u$ of dimensions $p \times 1$. The residual vector $\varepsilon$ is of dimensions $n \times 1$ and follows a multivariate normal distribution $N(0, \sigma_\varepsilon^2 I)$ with spherical covariance. The random effects $Zu$ are also modelled with a multivariate normal distribution $N(0, \sigma_g^2 K)$, where $K = ZZ^T$ is the Genetic Relatedness Matrix (GRM).

In FlexLMM the GRM is computed in a Leave-One-Chromosome-Out (LOCO) fashion, so as to avoid double-fitting the same variants as both fixed and random effects. The variance components $\sigma_\varepsilon^2$ and $\sigma_g^2$ are estimated using the `lmm.aireml` function of the gaston package, using only non-genetic fixed effects (more specifically, the design matrix corresponding to the user-defined null model is used). The residual variance-covariance matrix is then evaluated as $\Sigma = \sigma_g^2 K + \sigma_\varepsilon^2 I$, and regressed out from the phenotype vector and the design matrix of fixed effects using a Cholesky decomposition $\Sigma = LL^T$. The transformed design matrix $X_{mm} = L^{-1}X$ and response vector $y_{mm} = L^{-1}y$ can then be used in a simple linear regression to estimate fixed effects. Fixed effects are estimated separately for the null model design matrix that does not include any genetic variable, and then for the model including the genetic variant of interest (or a function of it). This second model is also specified by the user. The model including the genetic variant of interest is fit as many times as there are genetic variants to be tested, using always the same variance components for the decorrelation step that were beforehand estimated using the null model. This has the advantage of requiring only one fit for the variance components per chromosome, and the same variance components can then be used to regress out the correlation structure for all the genetic variant-specific models in the chromosome.

The use of user-defined models, that can be specified as pipeline parameters using the R formula interface, allows for maximum flexibility. The formulas can use a special term x, which is understood by the pipeline to refer to the current genetic variant of interest. This allows for fitting more complex models that can include non-linearities, such as a dominance term. Statistical significance is determined by performing a likelihood ratio test between the null model and the complete model defined by the user.

To account for multiple testing, the pipeline can perform permutations of the design matrix rows after the decorrelation step. Heritability and the variance-covariance matrix are estimated on the original dataset in this case. Permuting after the decorrelation step makes the residuals from the model uncorrelated, and so exchangeable under the null hypothesis of no effect due to genetic variants. The permutations define an empirical null distribution for the *p*-values, that can then be used to set an empirical significance threshold that corrects for multiple testing. For each permutation, all the design matrices are permuted in the same way, conserving the correlation structure among genetic variants. The minimum *p*-value obtained in each permutation across the genome is stored, and the set of minimum *p*-values is used to define the null distribution for genome-wide significance.

# Conclusion

We have developed FlexLMM, a Nextflow pipeline that can run linear mixed models for genome-wide association studies. The key advantage of our implementation is the flexibility in the definition of the exact statistical model to be used, and the native implementation of permutations for the definition of an empirical significance threshold. These features will make the pipeline particularly useful for scientists working with structured non-human populations, such as multi-parental F2 crosses among inbred lines of model organisms or farm animals and plants. The implementation using the Nextflow language allows for extreme portability to a variety of computational infrastructures, with minimal installation and tuning effort on the part of the user. Future additions to the software will include the implementation of tests for interaction terms, such as Gene by Environment interactions (GxE), and epistatic interactions (GxG). We also plan to add support for binary phenotypes.

# References

Alamin, M., Sultana, M. H., Lou, X., Jin, W., & Xu, H. (2022). Dissecting Complex Traits Using Omics Data: A Review on the Linear Mixed Models and Their Application in GWAS. *Plants*, *11*(23), Article 23. https://doi.org/10.3390/plants11233277

Chang, C. C., Chow, C. C., Tellier, L. C., Vattikuti, S., Purcell, S. M., & Lee, J. J. (2015). Second-generation PLINK: Rising to the challenge of larger and richer datasets. *GigaScience*, *4*(1), s13742-015-0047–0048. https://doi.org/10.1186/s13742-015-0047-8

Claire, D.-R., & Hervé, P. (2018). 46th European Mathematical Genetics Meeting (EMGM) 2018, Cagliari, Italy, April 18-20, 2018: Abstracts. *Human Heredity*, *83*(1), 1–29. https://doi.org/10.1159/000488519

Di Tommaso, P., Chatzou, M., Floden, E. W., Barja, P. P., Palumbo, E., & Notredame, C. (2017). Nextflow enables reproducible computational workflows. *Nature Biotechnology*, *35*(4), 316–319. https://doi.org/10.1038/nbt.3820

Gu, Z., Eils, R., & Schlesner, M. (2016). Complex heatmaps reveal patterns and correlations in multidimensional genomic data. *Bioinformatics*, *32*(18), 2847–2849. https://doi.org/10.1093/bioinformatics/btw313

Jiang, L., Zheng, Z., Qi, T., Kemper, K. E., Wray, N. R., Visscher, P. M., & Yang, J. (2019). A resource-efficient tool for mixed model association analysis of large-scale data. *Nature Genetics*, *51*(12), 1749–1755. https://doi.org/10.1038/s41588-019-0530-8

John, M., Ankenbrand, M. J., Artmann, C., Freudenthal, J. A., Korte, A., & Grimm, D. G. (2022). Efficient permutation-based genome-wide association studies for normal and skewed phenotypic distributions. *Bioinformatics*, *38*(Supplement_2), ii5–ii12. https://doi.org/10.1093/bioinformatics/btac455

John, M., Korte, A., & Grimm, D. G. (2023). *permGWAS2: Enhanced and Accelerated Permutation-based Genome-Wide Association Studies* (p. 2023.11.28.569016). bioRxiv. https://doi.org/10.1101/2023.11.28.569016

Joo, J. W. J., Hormozdiari, F., Han, B., & Eskin, E. (2016). Multiple testing correction in linear mixed models. *Genome Biology*, *17*(1), 62. https://doi.org/10.1186/s13059-016-0903-6

Loh, P.-R., Tucker, G., Bulik-Sullivan, B. K., Vilhjálmsson, B. J., Finucane, H. K., Salem, R. M., Chasman, D. I., Ridker, P. M., Neale, B. M., Berger, B., Patterson, N., & Price, A. L. (2015). Efficient Bayesian mixed-model analysis increases association power in large cohorts. *Nature Genetics*, *47*(3), 284–290. https://doi.org/10.1038/ng.3190

Mbatchou, J., Barnard, L., Backman, J., Marcketta, A., Kosmicki, J. A., Ziyatdinov, A., Benner, C., O'Dushlaine, C., Barber, M., Boutkov, B., Habegger, L., Ferreira, M., Baras, A., Reid, J., Abecasis, G., Maxwell, E., & Marchini, J. (2021). Computationally efficient whole-genome regression for quantitative and binary traits. *Nature Genetics*, *53*(7), 1097–1103. https://doi.org/10.1038/s41588-021-00870-7

R Core Team. (2023). *R: A Language and Environment for Statistical Computing*. R Foundation for Statistical Computing. https://www.R-project.org/

Scott, M. F., Fradgley, N., Bentley, A. R., Brabbs, T., Corke, F., Gardner, K. A., Horsnell, R., Howell, P., Ladejobi, O., Mackay, I. J., Mott, R., & Cockram, J. (2021). Limited haplotype diversity underlies polygenic trait architecture across 70 years of wheat breeding. *Genome Biology*, *22*(1), 137. https://doi.org/10.1186/s13059-021-02354-7

Sudlow, C., Gallacher, J., Allen, N., Beral, V., Burton, P., Danesh, J., Downey, P., Elliott, P., Green, J., Landray, M., Liu, B., Matthews, P., Ong, G., Pell, J., Silman, A., Young, A., Sprosen, T., Peakman, T., & Collins, R. (2015). UK Biobank: An Open Access Resource for Identifying the Causes of a Wide Range of Complex Diseases of Middle and Old Age. *PLOS Medicine*, *12*(3), e1001779. https://doi.org/10.1371/journal.pmed.1001779

Wickham, H. (2016). *ggplot2: Elegant Graphics for Data Analysis*. Springer-Verlag New York. https://ggplot2.tidyverse.org

Wilke, C. O. (2020). *cowplot: Streamlined Plot Theme and Plot Annotations for 'ggplot2'*.

https://CRAN.R-project.org/package=cowplot

Yang, J., Lee, S. H., Goddard, M. E., & Visscher, P. M. (2011). GCTA: A Tool for Genome-wide Complex Trait Analysis. *The American Journal of Human Genetics*, *88*(1), 76–82. https://doi.org/10.1016/j.ajhg.2010.11.011

Yang, J., Zaitlen, N. A., Goddard, M. E., Visscher, P. M., & Price, A. L. (2014). Advantages and pitfalls in the application of mixed-model association methods. *Nature Genetics*, *46*(2), 100–106. https://doi.org/10.1038/ng.2876

Zhou, W., Nielsen, J. B., Fritsche, L. G., Dey, R., Gabrielsen, M. E., Wolford, B. N., LeFaive, J., VandeHaar, P., Gagliano, S. A., Gifford, A., Bastarache, L. A., Wei, W.-Q., Denny, J. C., Lin, M., Hveem, K., Kang, H. M., Abecasis, G. R., Willer, C. J., & Lee, S. (2018). Efficiently controlling for case-control imbalance and sample relatedness in large-scale genetic association studies. *Nature Genetics*, *50*(9), 1335–1341. https://doi.org/10.1038/s41588-018-0184-y